**Controlling and distinguishing electronic transport of topological and trivial surface states in a topological insulator**

Helin Cao[1, 2, *, #], Chang Liu[3, 4], Jifa Tian[1, 2], Yang Xu[1, 2], Ireneusz Miotkowski[1], M. Zahid Hasan[3, 4], Yong P. Chen[1, 2, 5, *]

[1]Department of Physics and Astronomy, Purdue University, West Lafayette, IN 47907 USA

[2]Birck Nanotechnology Center, Purdue University, West Lafayette, IN 47907 USA

[3]Joseph Henry Laboratories, Department of Physics, Princeton University, Princeton, New Jersey 08544, USA

[4]Princeton Institute for Science and Technology of Materials, Princeton University, Princeton, New Jersey 08544, USA

[5]School of Electrical and Computer Engineering, Purdue University, West Lafayette, IN 47907 USA

[#]Current address: Department of Physics, University of Washington, Seattle, WA 98195, USA

*To whom correspondence should be addressed: helincao@gmail.com, yongchen@purdue.edu

**Topological insulators (TI), with characteristic Dirac-fermion topological surface states (TSS), have emerged as a new class of electronic materials with rich potentials for both novel physics and device applications[1–4]. However, a major challenge with realistic TI materials is to access, distinguish and manipulate the electronic transport of TSS often obscured by other possible parallel conduction channels that include the bulk as well as a two-dimensional electron gas (2DEG) formed near the surface due to bending of the bulk bands[5–9]. Such a (Schrodinger-fermion) 2DEG represents topologically-trivial surface states, whose coexistence with the TSS has been revealed by angle resolved photoemission spectroscopy[5–9]. Here we show that simple manipulations of surface conditions can be used to access and control both types of surface states and their coexistence in bulk-insulating $Bi_2Te_2Se$, whose surface conduction is prominently manifested in temperature dependent resistance and nonlocal transport. The trivial 2DEG and TSS can both exhibit clear**

**Shubnikov–de Haas oscillations in magnetoresistance, with different Berry phases ~0 and ~π that distinguish their different topological characters. We also report a deviation from the typical weak antilocalization behavior, possibly due to high mobility TSS. Our study enables distinguishing, controlling and harnessing electronic transport of TI surface carriers with different topological natures.**

Our high quality $Bi_2Te_2Se$ (BTS221) single crystals were synthesized by the Bridgman technique (see Method for the growth procedure, and Fig. S1 for an X-ray diffraction characterization), similar to those previously reported[10,11]. The TI nature of our crystals was verified by ARPES measurements. Figure 1a and b show the band structure maps of energy-momentum (*E-k*) dispersion of sample A obtained on an *in-situ* freshly cleaved (111) surface and an "aged" surface (subject to slightly degraded vacuum, see Method) respectively (see also Fig. S2 for the Fermi surface maps). The characteristic V-shaped (Dirac-like) TSS band was directly revealed on both fresh and aged surfaces. A prominent shift of band energies on the aged surface indicates a downward band bending, which n-dopes the surface. Despite the Fermi level shifting, the overall shape of TSS remained unchanged. The Fermi velocities ($v_F$) for the TSS extracted from the slopes of the *E-k* dispersions were both ~$5.5\times10^5$ m/s, comparable to previous ARPES measurements[12].

To probe surface conduction in our samples, we performed non-local transport measurements[13–15]. The lower inset of Fig. 2a depicts the measurement geometry. The lowercase letters label 6 contact leads on the top and bottom (111) surfaces (see Method). Figure 2a shows the temperature (T) dependence of the non-local voltage $V_{NL}$ (divided by an excitation current $I_{e,c}$) across different contact leads (e.g., $V_{f,b}$ denoting $V_{NL}$ between "f" and "b") measured in sample B (~160 μm thick). At high T (> ~65 K), both $V_{f,b}$ and $V_{f,g}$ were positive. On the other hand, $V_{a,b}$ is negative (the right inset of Fig. 2a) and approximately equals to $-V_{f,g}$, indicating that $I_{e,c}$ flew largely through the conductive bulk along the c-axis. As T decreased, the amplitudes of all the $V_{NL}$ increased by about 3 orders of magnitude, and the sign of $V_{a,b}$ also changed to positive. This suggests that a prominent portion of $I_{e,c}$ flew along the sample surfaces as schematically illustrated by the arrows in the left inset of Fig. 2a.

Another way to probe surface transport is to compare how different surface conditions influence the sample's transport properties. Figure 2b shows $T$-dependent resistivities $\rho_{xx}(T)$ obtained from standard (local) transport measurements in Sample "C" (~ 220μm thick, in a quasi-Hall-bar geometry as shown in Fig. 2b inset) with 5 different surface conditions (which we will refer to as Samples $C_0$, $C_0'$, $C_1$, $C_2$ and $C_3$, see details in supplementary Table S1). While "fresh" top surfaces ($C_0$ and $C_2$) were obtained by cleaving the sample with adhesive tapes followed by minimal delay and ambient exposure (no more than 1-2 min) before loading into an evacuated cryostat for transport measurements, various "aged" ones ($C_0'$, $C_1$ and $C_3$) were obtained after longer (accumulated) ambient exposure time (since the last fresh cleavage). Even though insulating behaviors ($\rho_{xx}$ increasing with decreasing $T$, attributed to an activated bulk) were always observed at high $T$, samples with fresh top surfaces ($C_0$ and $C_2$) were more conductive than those with aged ones ($C_0'$, $C_1$, $C_3$) at low $T$, where $\rho_{xx}$ starts saturating (consistent with remnant metallic surface conduction after the bulk conduction freezes out). At such low $T$ (< 20 K), besides the resistivity (Fig. 2b) measured at zero magnetic field (B), the magneto-resistance is also highly dependent on the surface conditions, as seen in Fig. 2c (showing resistance $R_{xx}$ of $C_0$, $C_1$ and $C_2$ as functions of B perpendicular to the cleaved surface). Furthermore, for each sample in Fig. 2c, its $R_{xx}$ under tilted B field was mostly controlled by the perpendicular component ($B_\perp \equiv B \cdot \cos\theta$), as shown in Fig. 2d (for $C_0$) and Fig. S5 ($C_1$ and $C_2$), where the tilt angle $\theta$ is defined between $\vec{B}$ and c-axis (normal to the top surface) as schematically shown in the inset of Fig. 2d. The angle dependent results suggest that the carriers giving rise to the magneto-resistances have a 2D nature, as expected for surface conduction. We notice a small deviation from 2D behavior can be observed at high $\theta$ and large B, possibly due to spin degrees of freedom[16] or bulk contributions.

It is important to point out that, while the nonlocal conduction (Fig.2a), the surface-sensitive resistance (Fig. 2b, c), and the 2D magneto-resistance (Fig. 2d) all indicate surface conduction, such a surface conduction does not necessarily arise (only) from the TSS, but could also arise from the trivial 2DEG, or both. Below we discuss how to distinguish these two types of surface carriers through quantum oscillations in the magnetoresistance. Figure 3a, b and c display the magnetoresistance (after subtracting polynomial backgrounds), $\Delta R_{xx}(\frac{1}{B_\perp})$, measured for various tilt angles $\theta$, in $C_0$, $C_1$ and $C_2$ respectively, where oscillatory patterns are clearly observed. Fourier analyses of the oscillations measured at $\theta = 0^o$ are shown in the right insets of Fig. 3d-f. Single frequencies ($B_F$) of 51.5 $T$ and 52.4 $T$ are extracted from $C_0$ and $C_1$ respectively. Remarkably, in $C_2$, we clearly resolve *two* different sets of oscillations (labeled as $C_2^I$ and $C_2^{II}$), with distinct frequencies of 53 T ($C_2^I$) and 254 T ($C_2^{II}$) respectively. All the oscillations in the three samples are periodical in 1/B, and are interpreted as Shubnikov-de Haas oscillations (SdHO) due to the formation of Landau levels (LL) in high magnetic fields. Similar to the overall magneto-resistance (Fig. 2d and Fig. S5), the SdHO (the positions, frequencies and even amplitudes) were also found to be mostly controlled by $B_\perp$ (with oscillations no longer observable within our B field range when $\theta$ approaches 90º), indicating a 2D nature consistent with surface carriers.

Even though both topological (TSS) and trivial (2DEG) surface states could give rise to SdHO, cyclotron orbits of the massless Dirac fermions in TSS would acquire a Berry phase of $\pi$, whereas those of Schrödinger fermions in the trivial 2DEG would have a 0 Berry phase. The Berry phase can be experimentally extracted from SdHO using the LL fan diagram. For this purpose, we should assign the LL indices (N) to *minima* in $G_{xx}$,[17,18] where $G_{xx} \equiv \frac{R_{xx}}{R_{xx}^2 + R_{xy}^2}$ denotes the diagonal element of the conductance tensor. Since $R_{xx}$ of our samples is at least ~2 orders of magnitudes larger than the Hall resistance ($R_{xy}$), minima in $G_{xx}$ correspond to *maxima* in $R_{xx}$ ($\Delta R_{xx}$, Supplemental Fig. S7), which we label by the dashed lines and integer LL indices in Fig. 3a-c. Figure 3d, e and f show LL fan diagrams, where $B_N$ is the field position of the $N^{th}$ LL, plotted for the $\theta=0^o$ SdHO in $C_0$, $C_1$ and $C_2$, respectively. To reduce ambiguities, we fixed the slope ($B_F$, obtained from the Fourier analyses), and used the y-intercept ($\beta$) as the only fitting parameter. The fitting in Fig. 3d yields $\beta = 0.42 \pm 0.02$ for $C_0$, giving a Berry phase ($2\pi\beta$) ~$0.8\pi$, close to $\pi$ for ideal Dirac fermions. In contrast, $\beta$ for $C_1$ was extracted to be $0 \pm 0.03$ (Fig. 3e), with Berry phase ~0, suggesting the observed SdHO arise from a topologically trivial 2DEG (Schrodinger fermions). Intriguingly, the linear fittings in Fig. 3f yield y-intercepts of $0.05 \pm 0.02$ for $C_2^I$ (Berry phase ~0) and $0.41 \pm 0.05$ for $C_2^{II}$ (Berry phase ~$0.8\pi$), indicating that they arise from a trivial 2DEG ($C_2^I$) and TSS ($C_2^{II}$) respectively, *coexisting* in Sample $C_2$.

To obtain more information about the different types of surface carriers, we analyze the temperature dependence of SdHO by Lifshitz-Kosevich theory[19]: $\frac{\Delta R_{xx}}{R_0} \propto \frac{2\pi^2 k_B T/\hbar\omega_c}{\sinh(2\pi^2 k_B T/\hbar\omega_c)}$, where $k_B$ is the Boltzmann's constant. The cyclotron frequency $\omega_c$ is related to the cyclotron mass (effective mass) as $\omega_c = eB/m_c$ (see Fig. S8 for detailed analysis, and Table 1 for extracted parameters). The cyclotron masses $m_c$ extracted from $C_0$ and $C_2^{II}$ are $0.08 m_e$ and $0.18 m_e$ ($m_e$ is the free electron mass) respectively. The Fermi velocity $v_F = \hbar k_F/m_c$ was found to be ~$6 \times 10^5$ $m/s$ for both $C_0$ and $C_2^{II}$, consistent with our and previous measurements for TSS in BTS221[10–12]. The constant Fermi velocity and linear relation $m_c \propto k_F$ (see Table 1) provide direct evidence for the Dirac fermion dispersion[20–22], further confirming the TSS origin of $C_0$ and $C_2^{II}$. On the other hand, $m_c$ was extracted to be $0.37\ m_e$ for $C_2^I$ (attributed to trivial 2DEG arising from the bulk conduction band) with a much smaller $v_F$~$1.3 \times 10^5$ $m/s$.

Given that the observed SdHO were highly sensitive to the top surface conditions, we attribute the SdHO to the carriers from the *top* surface, where the bottom surface (in direct contact to the substrate with an adhesive polymer) is believed to have substantially lower mobility and did not give rise to observable SdHO. The different surface transport and SdHO observed in the same sample "C" after different exposure ("aging") process can be further understood as follows. In the "fresh" $C_0$ (exposure time < 30 secs), the surface Fermi level ($E_F = \frac{\hbar^2 k_F^2}{m_c}$)[23] is 148 meV above the Dirac point ($E_D$), and is in the bulk band gap (as illustrated in the left inset of Fig. 3d), such that only the SdHO arising from the TSS (with an extracted quantum SdH mobility of 1800 cm$^2$/Vs, see Table 1) was observed. After a longer surface exposure (73 days), $E_F$ (now ~330 meV above $E_D$) of the "aged" $C_1$ was shifted above the bottom of the (downward-bent) conduction band near the surface, inducing a topologically-trivial surface 2DEG populating the triangular shaped quantum well (depicted in the left inset of Fig. 3e) that coexists with the TSS. However, the long exposure in ambient can introduce a substantial amount of surface impurities (scattering centers) that strongly scatter TSS carriers (residing at the surface), such that the mobility of TSS became too low to give observable SdHO in $C_1$. Only SdHO from the trivial 2DEG (located deeper below the surface[9] thus less affected by the surface impurities, with extracted quantum SdHO mobility ~1900 cm$^2$/Vs), was observed in $C_1$ within our B field range. After a re-cleavage followed by a short exposure (~ 2 min) for $C_2$, while $E_F$ remains similar as that in $C_1$ (seen from the similar 2DEG carrier densities extracted from $C_1$ and $C_2^I$), the quantum mobility of TSS presumably became high enough (extracted to be ~300 cm$^2$/Vs from $C_2^{II}$, Table 1) to give observable SdHO ($C_2^{II}$), coexisting with the SdHO ($C_2^I$) of the trivial 2DEG (with extracted quantum mobility ~2000 cm$^2$/Vs, Table 1). It is interesting to note that the topological-protection for TSS only prohibits elastic back scattering, whereas angled scattering (which limits the quantum mobility) is not prohibited and can be more significant for TSS (more prone to surface impurities) compared to the 2DEG (buried and better protected from the surface impurities), leading to *lower* quantum mobility observed for TSS than that for topological-trivial 2DEG (as in $C_1$, $C_2$, see Table 1).

From the Fermi energies (see Table 1) extracted from the two sets of SdHO ($C_2^I$ and $C_2^{II}$), we can further deduce the energy separation between the Dirac point and the bottom of the bulk conduction band to be 313meV, in reasonable agreement with the ARPES measurements (Fig. 1 and Ref. 12).

Finally, we discuss magnetoresistance measured in low B field. Figure 4a shows $\Delta G_{xx} \equiv G_{xx} - G_0$, where $G_0 = G_{xx}(B = 0T)$, of $C_3$ (exposed for 13 days to obtain an aged surface) measured at various temperatures. An overall negative magneto-conductance, weakened at elevated temperatures, can be attributed to the weak anti-localization (WAL) as a result of destructive quantum interference between time-reversed trajectories in a diffusive transport. The WAL in $C_3$ can be fitted (the black dashed lines) by the Hikami-Larkin-Nagaoka (HLN) theory[24]: $\Delta G_{xx}(B) = -\alpha \frac{e^2}{2\pi^2 \hbar}\left[\ln\left(\frac{\hbar}{4el_\phi^2 B}\right) - \Psi\left(\frac{\hbar}{4el_\phi^2 B} + \frac{1}{2}\right)\right]$, with $\Psi$ being the Digamma function. Figure 4b shows the two extracted fitting parameters, $l_\phi$ (phase coherent length) and $\alpha$ (a prefactor, which is $-\frac{1}{2}$ for each coherent diffusive transport channel), as functions of T. A strong decay of $l_\phi$ at increased T was found to follow a power-law dependence with a power $\sim -0.32 \pm 0.07$. For an ideal TI, the prefactor $\alpha$ is expected to be -1, indicating two transport channels from decoupled top and bottom surfaces. We extract $\alpha$ in $C_3$ to range from $\sim$ -7 to -12, suggesting that carriers from other transport channels, such as the surface 2DEG or remnant conducting regions in the bulk, may still contribute to the observed WAL. We note both TSS Dirac fermions (with the Berry phase $\pi$ associated with spin-momentum locking)[25] and the Schrödinger fermions originated from the bulk[26,27] (with strong spin-orbit coupling) can give destructive quantum interferences that lead to WAL.

Interestingly, such a good fitting of the observed WAL behavior to the HLN theory as demonstrated above was only observed in samples with an aged top surface (such as $C_3$ above). In Fig. 4c, we plot normalized magneto-conductance $G_{xx}/G_0$ measured in samples with four different surface conditions. For each sample, a sharp peak (occurring at magnetic field $B_1$ labeled in Fig. 4d) is observed in the derivative of magneto-conductance, $-\frac{dG_{xx}}{dB}$. This $B_1$ is determined by the dephasing field $\frac{\hbar}{4el_\phi^2}$ according to the HLN theory. For B > $B_1$, $-\frac{dG_{xx}}{dB}$ of $C_1$ and $C_3$ (aged top surface) monotonously decreases with B. The $-\frac{dG_{xx}}{dB}$ for samples $C_1$ and $C_3$ can be well fitted to the HLN theory (the black dashed lines in Fig. 4d, also see supplementary Fig. S10). However, in $C_0$ and $C_2$ (fresh surface giving rise to SdHO of TSS), we observed another peak (labeled by $B_2$ and much broader than the $B_1$ peak) in $-\frac{dG_{xx}}{dB}$ not captured by the standard WAL behavior as described by HLN theory. Curiously, the $B_2$ peak did not vanish at higher T as the $B_1$ peak did (supplementary Fig. S11). The origin of this low-field magnetoconductance feature ($B_2$ peak), only seen with fresh-cleaved surfaces with relatively high mobility TSS, remains to be better understood. We speculate that such a feature may relate to the high mobility TSS and the interplay between such TSS with other transport channels.

In conclusion, our scheme of manipulating the surface condition enabled us to access transport signatures arising from surface carriers with different topological natures in 3D topological insulators, leading to a demonstration of the co-existence of TSS and the surface 2DEG through quantum oscillations in high magnetic fields. We also studied surface conduction in zero and low B fields with different surface conditions. Our findings pave the way to distinguishing, controlling and harnessing the two distinct types of surface carriers for device applications and fundamental research.

**Methods**

*Crystal growth.* The single crystals of $Bi_2Te_2Se$ (BTS221) are grown by the Bridgman method using a two-step procedure. In the first step, the raw material is synthesized from high purity elements in a two-zone horizontal furnace with independent temperature control. The starting materials, 6N (99.9999%) purity Bi, Te and Se, are deoxidized before use. Both Te and Se are purified further by multiple vacuum distillations under dynamic vacuum of $10^{-7}$ torr. The synthesis is made in vitreous carbon boats to avoid possible contamination from quartz. After the synthesis is completed at 900 °C, the pre-reacted charge is slowly cooled down under a controlled pressure of Se. The charge is then transferred into a carbonized quartz ampoule. The growth ampoule is placed in a vertical Bridgman three-zone furnace with independent temperature control. The linear gradient in the growth zone is set to 12 °C/cm. The radial gradient inside the growth zone is symmetric and estimated to be less than 0.5 °C/cm. The typical speed for moving the ampoule through the temperature gradient is ~1 mm/hour. The crystal structure of the as-grown crystal was examined by X-ray powder diffraction (XRD), and a representative pattern is shown in supplementary Fig. S1.

*ARPES measurements.* BTS221 crystals with sufficient n-type doping were selected in favor of ARPES. The measurements were performed at the APPLE-PGM beam line of the Synchrotron Radiation Center (SRC), Wisconsin, equipped with a VG-Scienta 200U electron analyzer. Samples were measured at ~11 K in a vacuum less than $5\times10^{-11}$ torr with 18 eV incident photon energy. Energy resolution was set to be ~20 meV. To obtain an "aged surface", we transferred the sample to another area (with a slightly higher residual gas pressure of ~$5\times10^{-9}$ torr) inside the vacuum chamber, and kept it there for ~1 hour.

*Device fabrication and transport measurements.* We isolated small pieces of single crystals with razor blades, and fabricated them into different types of devices using silver paint or indium as electrodes. Transport measurements were performed in a He-3 system at the National High Magnetic Field Laboratory (NHMFL), and a variable temperature insert (VTI) system at Purdue University. The transport data shown in this paper were measured using standard low frequency (~10 Hz) lock-in technique (Stanford Research System 830). Similar results were also obtained by DC measurements (with a Keithley 2400 Sourcemeter). For samples used in magneto-transport measurements, their bottom surfaces were glued onto chip carriers with GE varnish. In the non-local measurements, $R_{NL}$ is confirmed to be independent of $I_{e,c}$ in a large current range. To ensure the voltage probes are far away from the path of conventional current flow, we made the spacing between any neighboring leads on the same surface at least ~2 mm, which is 15 times greater than the sample's thickness (~ 160 μm).

**Acknowledgement**

This work was mainly supported by DARPA MESO program (Grant N66001-11-1-4107). High magnetic field transport measurements were performed at the National High Magnetic Field Laboratory (NHMFL), which is jointly supported by the National Science Foundation (DMR0654118) and the State of Florida.

M.Z.H. acknowledges support from the A. P. Sloan Foundation. The ARPES measurements were performed at the APPLE-PGM beam line of the Synchrotron Radiation Center, Stoughton, Wisconsin. We thank E. Palm, L. Engel, J. J. Jaroszynski from NHMFL and P. Metcalf from Purdue for experimental assistance. The authors also thank Yuli Lyanda-Geller and Hsin Lin for valuable discussions.

**Figures:**

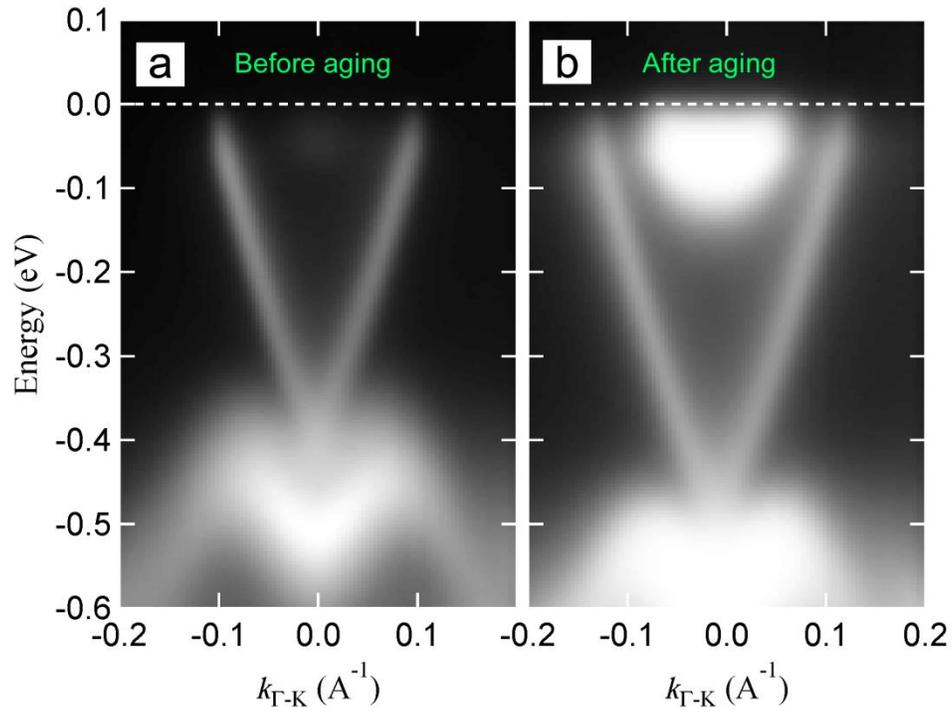

**Figure 1. ARPES-measured electronic band structure of Bi$_2$Te$_2$Se (BTS221), revealing topological surface states (TSS).** The *E-k* band map along the $\Gamma - K$ direction measured by ARPES (angle-resolved photoemission spectroscopy) on (a) a freshly cleaved sample ("Sample A") and (b) the same sample after an aging process. The Fermi level in each case is labeled by the horizontal dashed line. The Fermi surface maps are displayed in Supplementary Fig. S2.

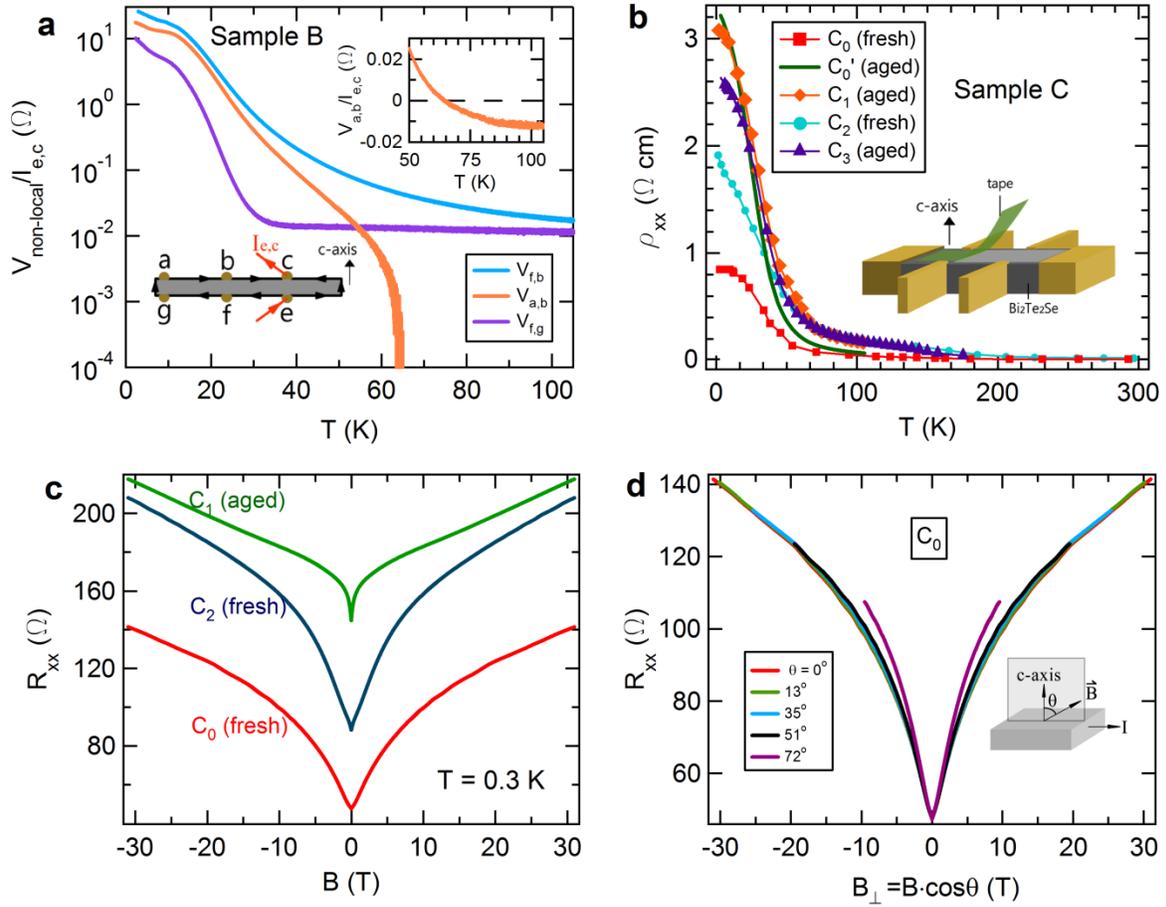

**Figure 2. Surface conduction revealed by non-local transport and surface-condition sensitive transport measurements.** (a) Temperature ($T$) dependence of non-local voltage $V_{NL}$ divided by $I_{e,c}$ measured in Sample B. The sample displayed an insulating behavior in the conventional $T$-dependent (local) resistivity measurement (see supplementary Fig. S3). Lower inset shows a schematic diagram of the measurement geometry (see Method and main text for details). Upper inset shows the sign of $V_{a,b}$ changing from negative to positive as $T$ decreases, indicating more current flowing on the surfaces as illustrated in the lower inset. (b) $T$-dependent resistivity $\rho_{xx}$ of sample C, which has been subjected to successive cleavages and air exposure to modify the top surface conditions as described in the text, all showing an insulating behavior. Below ~20 K, the surface conduction becomes prominent and the samples with a fresh top surface ($C_0$ and $C_2$) are more conductive than those with an aged top surface ($C_0$',

$C_1$ and $C_3$). (c) Magneto-resistance $R_{xx}$ of $C_0$, $C_1$ and $C_2$ measured in perpendicular magnetic field (B) at 0.3 K. (d) $R_{xx}$ plotted against $B_\perp = B \cdot \cos(\theta)$ measured on $C_0$ in tilted B. Inset illustrates the tilt angle ($\theta$) from perpendicular direction (c-axis). The $R_{xx}$ for all three samples (see Fig. S5 for similar data from $C_1$ and $C_2$) was found to be mostly controlled by $B_\perp$, suggesting a 2D transport behavior.

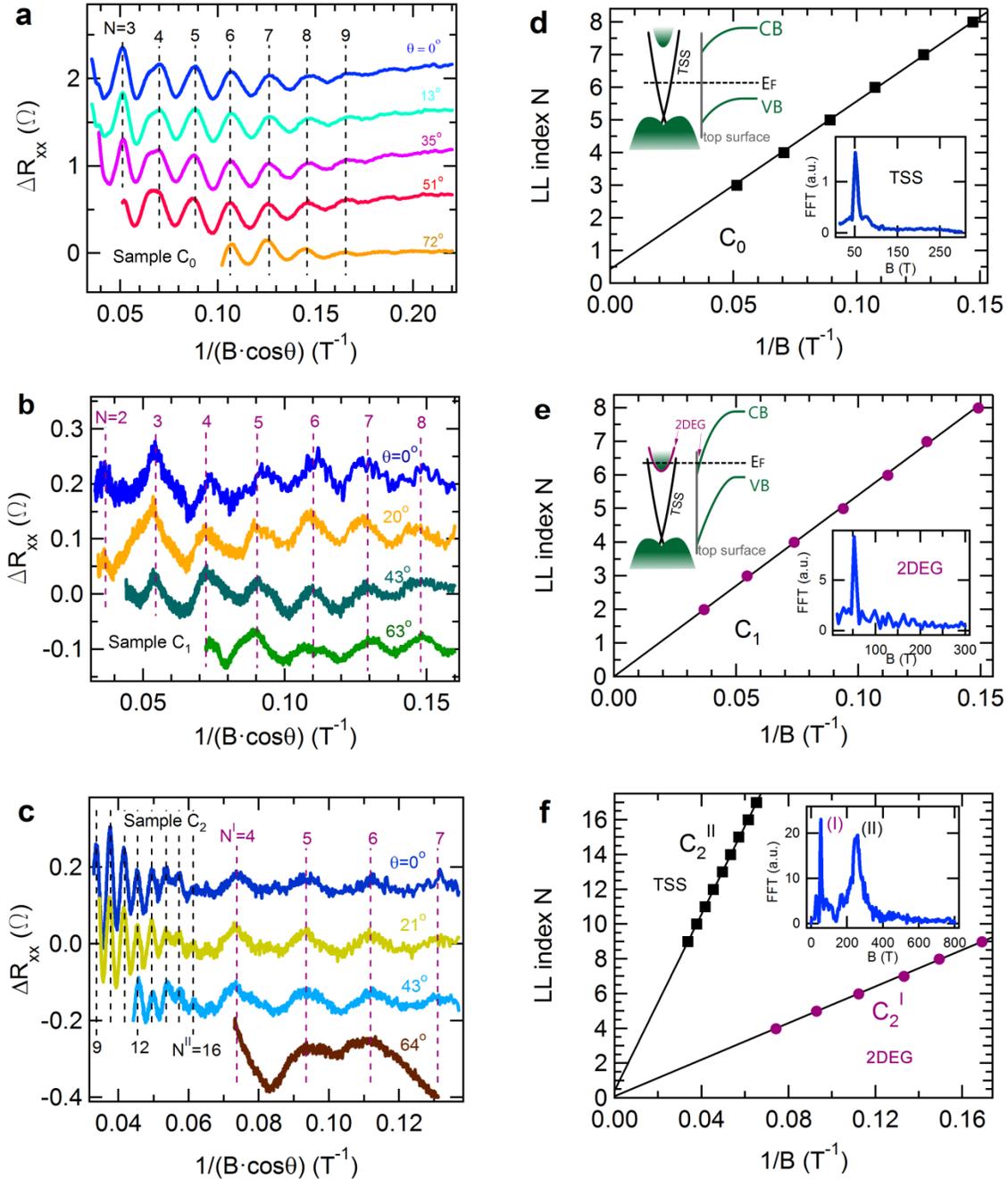

**Figure 3. Shubnikov–de Haas oscillations (SdHO) from TSS and/or trivial 2DEG.** Oscillatory component $\Delta R_{xx}$ ($R_{xx}$ with a smooth polynomial background subtracted), plotted against $1/B_\perp$ at various tilt angles (θ) measured in Sample C with three different surface conditions (a) $C_0$, (b) $C_1$ and (c) $C_2$ (curves shifted vertically for clarity). Selected Landau level (LL) indices (*N*) are labeled. The oscillations were found to be controlled by the perpendicular component of magnetic field ($B_\perp$), indicating 2D nature of the responsible charge carriers. (d, e, f) Landau fan diagram, *N* vs $\frac{1}{B_N}$ ($B_N$ denoting the B field positions for $R_{xx}$ maxima, corresponding to minima in conductance $G_{xx}$), for the *θ*=0 data (shown in a-c) for $C_0$, $C_1$ and $C_2$ respectively. Solid lines are linear fitting using the y-axis intercept as the only free parameter, while the slope was fixed by the SdHO frequencies obtained from Fourier analysis of $\Delta R_{xx}(\frac{1}{B})$ as displayed in the right insets of (d-f). The intercepts for $C_0$ and $C_2$(II) are ~0.4, consistent with Dirac fermions; and those for $C_1$ and $C_2$(I) ~0, consistent with Schrödinger fermions. Left insets of (d, e) schematically depict the band bending near surface for $C_0$ (d), and for $C_1$ and $C_2$ (e).

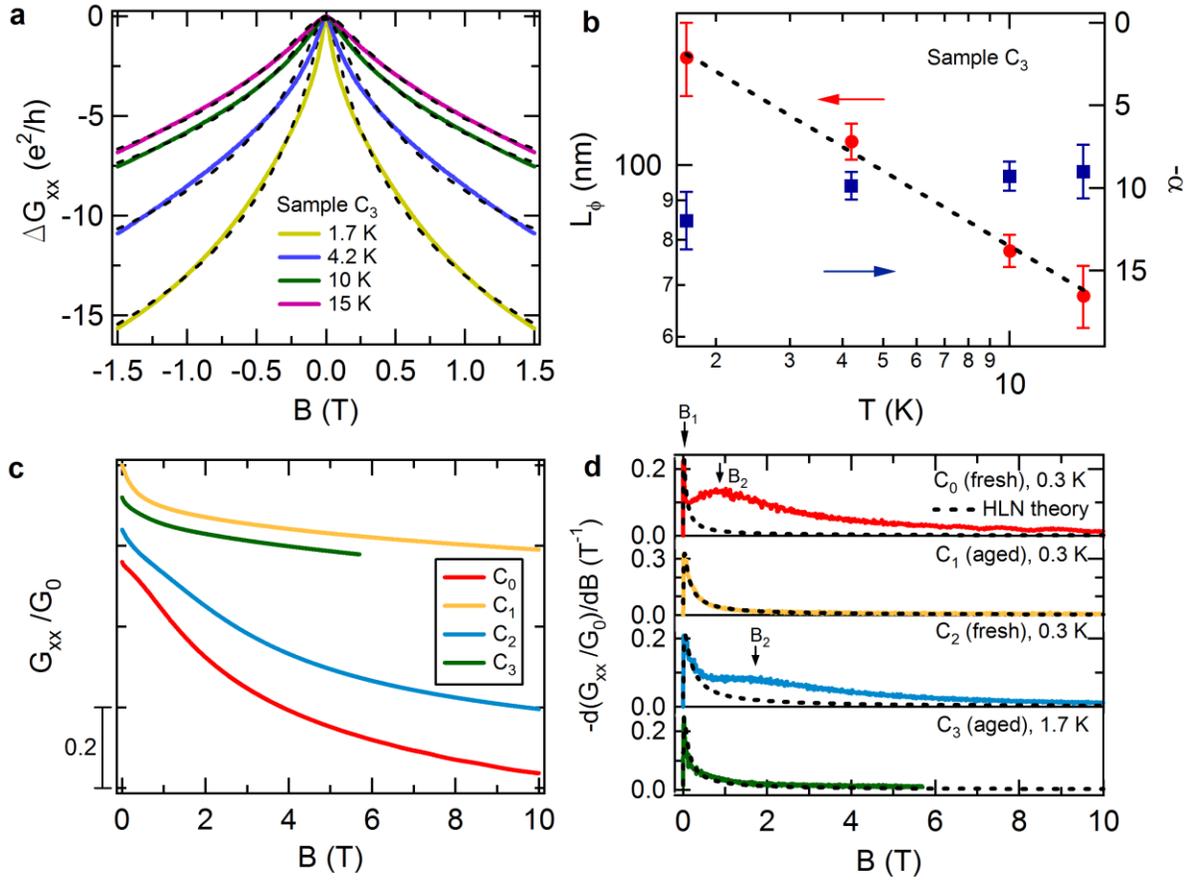

**Figure 4. Low B-field magnetotransport exhibiting or deviating from conventional weak anti-localization (WAL).** (a) Magnetoconductance $\Delta G_{xx} \equiv G_{xx} - G_0$, where $G_0 = G_{xx}(B = 0T)$, of $C_3$ at various temperatures. The peaks observed at B=0 T can be interpreted as due to the WAL. The data are fitted to HLN (Hikami-Larkin-Nagaoka) theory as shown by the dash lines. (b) Extracted prefactor $\alpha$ and phase coherence length $l_\phi$ from HLN fitting at various temperatures. The dashed line shows the power-law dependence of $l_\phi$ on T ($l_\phi \propto T^{-0.32 \pm 0.07}$, noting both $l_\phi$ and T are plotted on the log scale). (c, d) Normalized magneto-conductance $G_{xx}/G_0$ and corresponding derivatives $\frac{-d(G_{xx}/G_0)}{dB}$ as functions of B of $C_0$, $C_1$, $C_2$ and $C_3$ measured at low temperatures (indicated in d). The curves in (c) are shifted vertically for clarity. The sharp peak labeled as "$B_1$" for all the samples in (d) is attributed to WAL and consistent

with HLN fitting. An additional broad peak "$B_2$" is observed in the "fresh surface" samples $C_0$ and $C_2$. Black dashed lines in (d) are calculated with the HLN fitting (parameters listed in Table S2).

| Sample | exposure time | SdHO | $B_F$ (T) | $k_F$ ($10^8$ m$^{-1}$) | $n_{2D}^{SdH}$ ($10^{12}$ cm$^{-2}$) | $\beta$ | $m_c$ ($m_e$) | $v_F$ ($10^5$ m/s) | $\mu_S^{SdH}$ (cm$^2$/Vs) | $E_F$ (meV) |
|---|---|---|---|---|---|---|---|---|---|---|
| $C_0$ | < 30 secs | $C_0$ | 51.5 | 3.9 | 1.2 | 0.42 | 0.08 | 5.7 | 1800 | 148 |
| $C_1$ | 73 days | $C_1$ | 52.4 | 4.0 | 2.5 | 0 | - | - | ~2000 | - |
| $C_2$ | ~2 mins | $C_2^I$ | 53 | 4.0 | 2.5 | 0.05 | 0.37 | 1.3 | 2000 | 17 |
| | | $C_2^{II}$ | 254 | 8.7 | 6.0 | 0.41 | 0.18 | 5.7 | 300 | 330 |

**Table 1. Sample C with 3 different surface conditions and corresponding transport parameters extracted from the SdH oscillations (SdHO).** Exposure time refers to the lapse following the last fresh cleaved surface. $B_F$ is the SdHO frequency and used to calculate the Fermi momentum $k_F = \sqrt{2B_F e/\hbar}$. Carrier density $n_{2D}^{SdH} = \frac{k_F^2}{4\pi}$ for TSS, and $=\frac{k_F^2}{2\pi}$ for 2DEG due to the spin degeneracy. LL intercept $\beta$ (with Berry phase = $2\pi\beta$) is found from the LL fan diagram (Fig. 3d-f). Effective mass $m_c$ is extracted from the T dependence of SdHO (Fig. S8) and Fermi velocity $v_F = \hbar k_F/m_c$. $\mu_S^{SdH}$ is SdHO quantum mobility extracted from the B dependence of SdHO (Figs. S8, S9). Fermi level $E_F = \frac{\hbar^2 k_F^2}{m_c} = \hbar k_F v_F$ for TSS[23] (relative to the Dirac point as in $C_0$ and $C_2^{II}$), and $E_F = \frac{\hbar^2 k_F^2}{2m_c}$ for 2DEG[23] (relative to the bottom of the conduction band as in $C_2^I$). The difference in $E_F$ between $C_2^I$ and $C_2^{II}$, 313meV, measures the separation between the Dirac point and bottom of the bulk conduction band, and is comparable with the ARPES measurements (Fig. 1 and Ref. 12). Dash entries indicate parameters not directly measured.

SUPPLEMENTARY INFORMATION

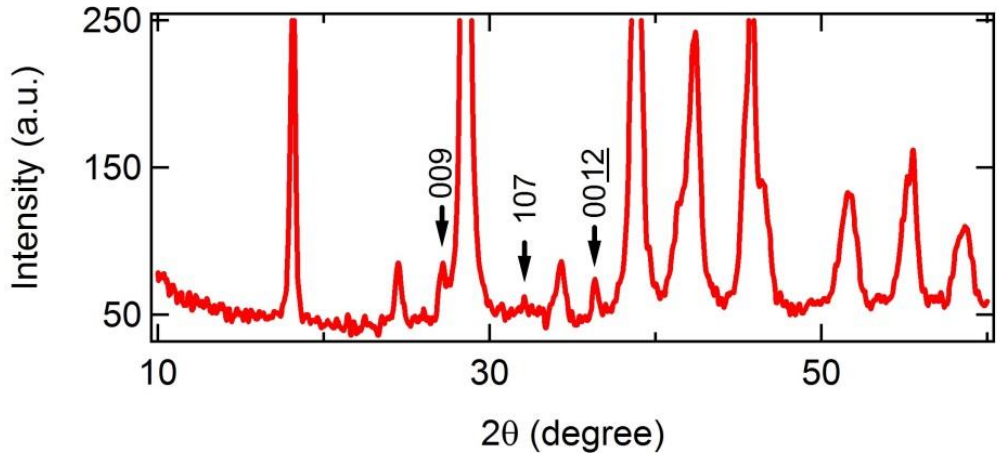

**Figure S1**. A representative X-ray powder diffraction (XRD) pattern on an as-grown $Bi_2Te_2Se$ (BTS221), showing three characteristic peaks, (009), (107) and (00$\underline{12}$), indicative of ordered Se and Te layers as previously reported[S1].

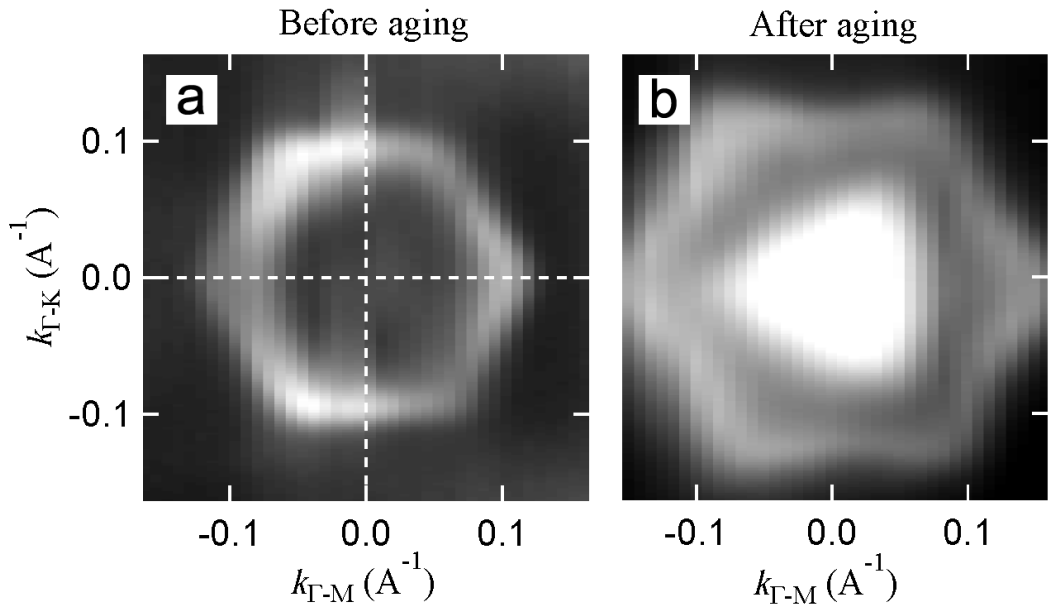

**Figure S2.** (a, b) ARPES Fermi surface (FS) maps corresponding to Fig. 1a and b respectively. The FS warping effect in the TSS is more pronounced after the aging process (which leads to higher Fermi energy

and surface carrier density). The triangular shape of the bulk conduction band (BCB) shown in (b) experiences a mirror reflection with respect to the $\Gamma - K$ axis as the incident photon energy changes from 18 eV to 22 eV (not shown), confirming that it originates from the bulk states.

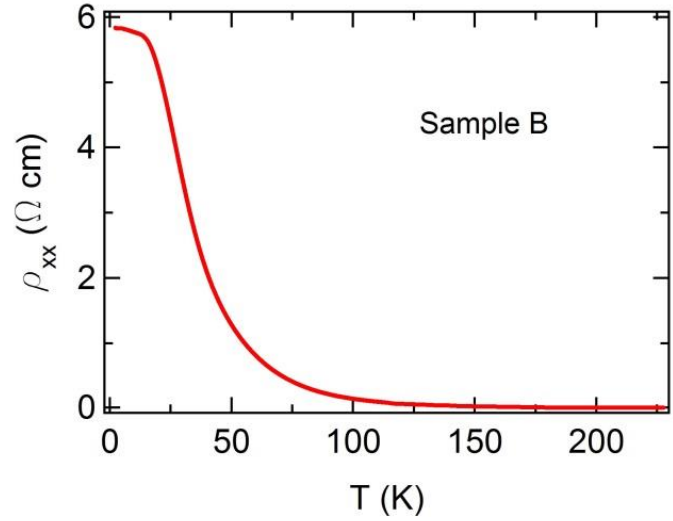

**Figure S3.** Conventional 4-terminal local resistivity ($\rho_{xx}$) of sample B (see maintext Fig. 2a) as a function of temperature (T) shows a bulk insulating behavior. The saturation in $\rho_{xx}$ at low T can be attributed to the metallic surface transport.

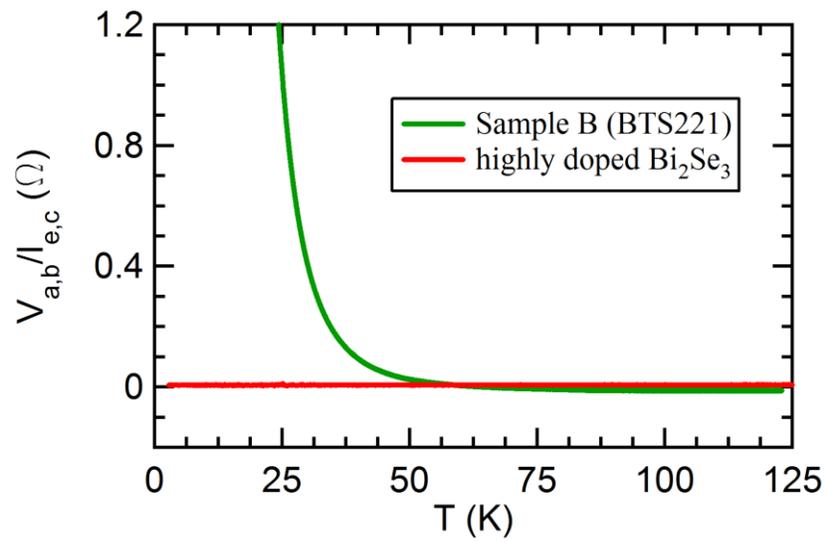

**Figure S4.** Comparison of the non-local resistance $V_{a,b}/I_{e,c}$ between sample B (BTS221), presented in Fig. 2a, and a highly doped (thus with a metallic bulk) $Bi_2Se_3$. Both samples have similar thickness and measurement geometry (see the inset of Fig.2a). The non-local resistance in the metallic $Bi_2Se_3$ sample remains ~0 down to 3 K.

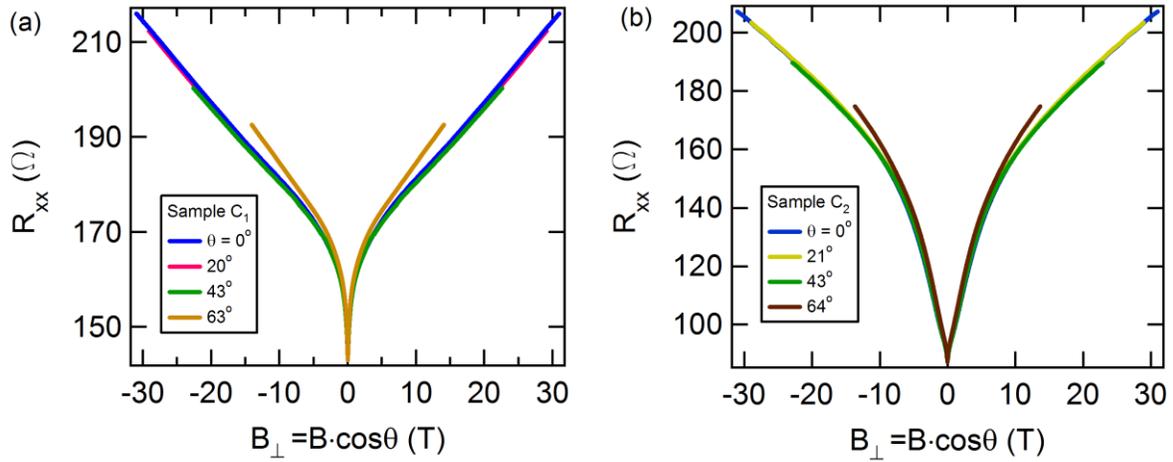

**Figure S5.** Similar to the measurements shown in Fig. 2d, magneto-resistance $R_{xx}$ of (a) $C_1$ and (b) $C_2$ is plotted against $B_\perp = B \cdot \cos(\theta)$ measured at 0.3K in tilted B with various tilting angle $\theta$, indicating 2D transport behaviors

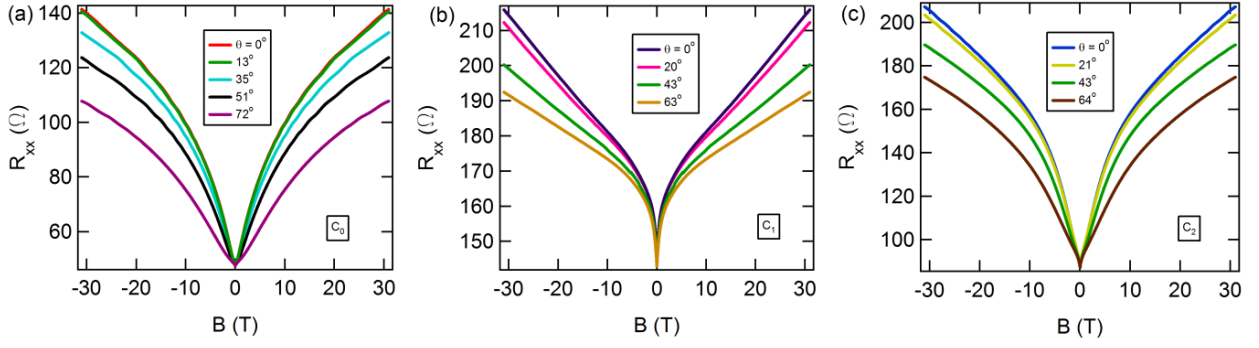

**Figure S6.** Four terminal magneto-resistance ($R_{xx}$, same data as shown in Fig. 2d, Fig. S5a, Fig. S5b, respectively) plotted against total B measured at 0.3 K for various tilt angles in (a) sample $C_0$, (b) $C_1$, and (c) $C_2$. In contrast to Fig. 2d, Fig. S5a, Fig. S5b, here the curves in each figure do not collapse.

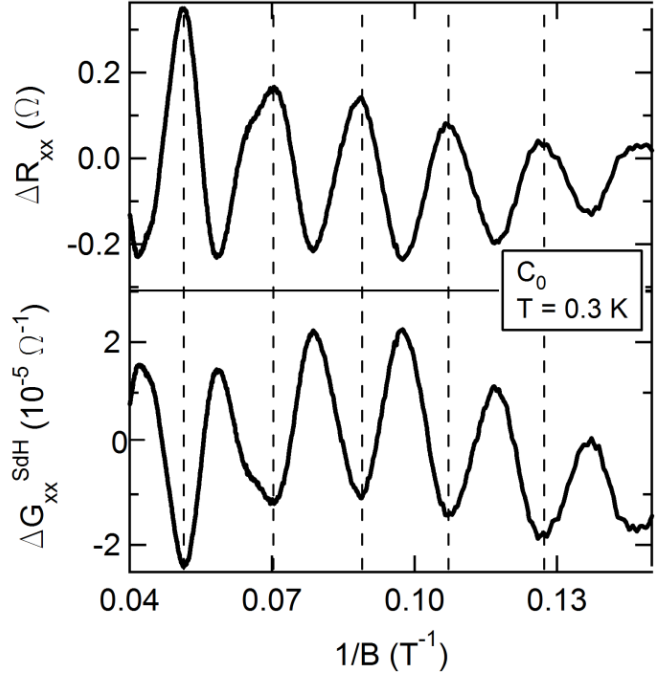

**Figure S7.** Oscillatory component $\Delta R_{xx}$ and $\Delta G_{xx}^{SdH}$ (obtained by subtracting polynomial backgrounds from $R_{xx}$ and $G_{xx} \equiv \frac{R_{xx}}{R_{xx}^2 + R_{xy}^2}$ respectively) as functions of 1/B measured in Sample $C_0$ at 0.3 K (main text Fig. 3a), showing that minima in $\Delta G_{xx}^{SdH}$ correspond to maxima in $\Delta R_{xx}$.

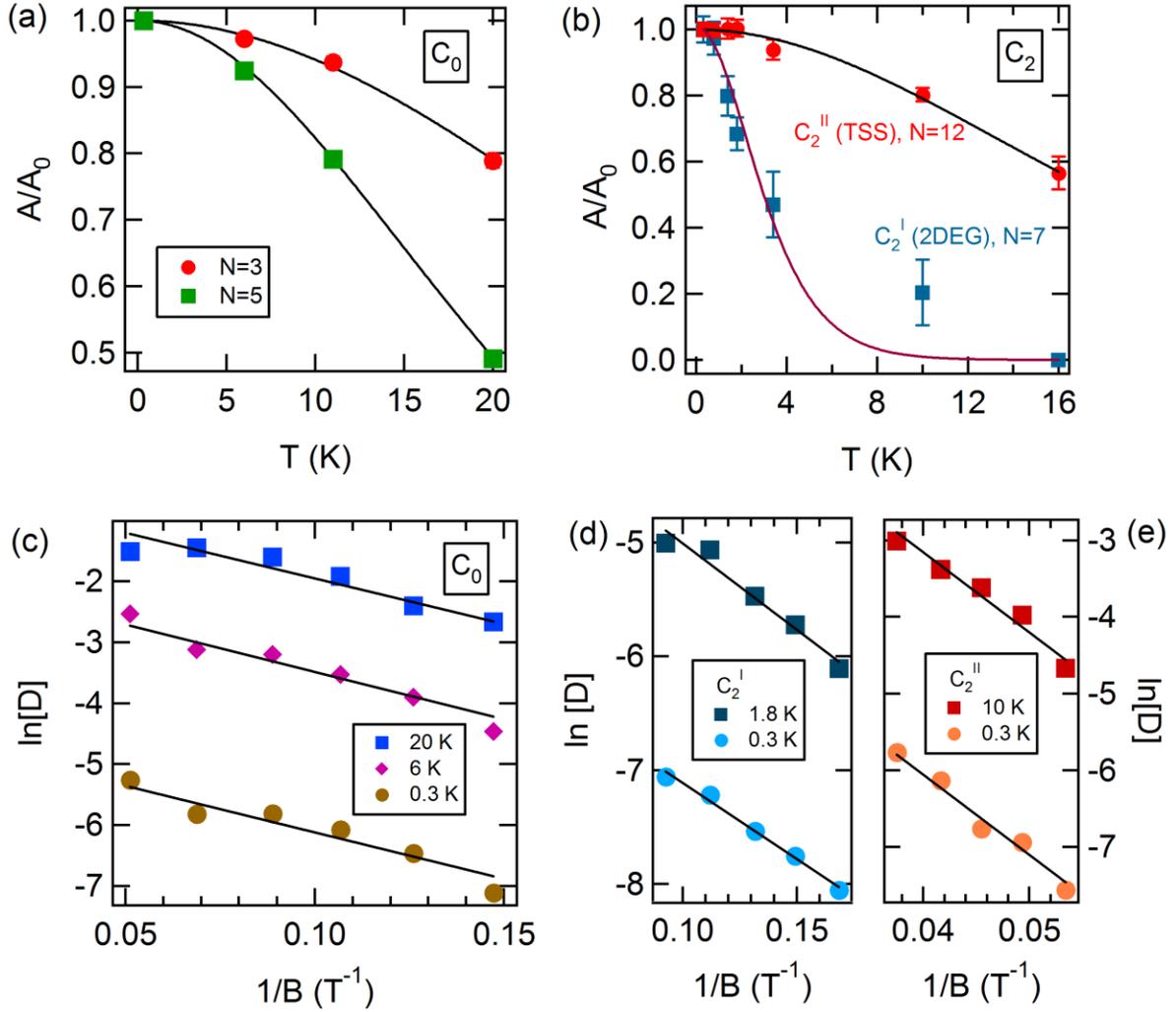

**Figure S8.** We use Lifshitz-Kosevich theory[S2] $\frac{\Delta R_{xx}(B,T)}{R_0} \propto \frac{\frac{2\pi^2 k_B T}{\hbar\omega_c}}{\sinh\left(\frac{2\pi^2 k_B T}{\hbar\omega_c}\right)} \cdot e^{-\frac{2\pi^2 k_B T_D}{\hbar\omega_c}}$ to analyze the temperature ($T$) and magnetic field ($B$) dependent SdH oscillation (SdHO) amplitudes (see relevant equations in the main text including the caption of Table 1). (a, b) Examples of the $T$ dependence of normalized SdHO amplitude (associated with representative Landan levels with indices N labeled), $\frac{A(T)}{A(T_0=0.3\,K)}$ for $C_0$, $C_2^I$ and $C2^{II}$. The solid lines are fittings to $\frac{A(T)}{A(T_0=0.3\,K)} = \frac{T}{\sinh(2\pi^2 k_B T m_c/\hbar eB)} \cdot \frac{\sinh(2\pi^2 k_B T_0 m_c/\hbar eB)}{T_0}$ with $m_c$ as the only fitting parameter. The fittings yield $m_c = 0.08 m_e$, $0.37 m_e$ and $0.18 m_e$ for $C_0$, $C_2^I$ and $C_2^{II}$, respectively. (c, d, e) Dingle plot of $\ln[D]$, where $D = (\Delta R/R_0)B\sinh(2\pi^2 k_B T m_c/\hbar eB)$, vs. $1/B$, at selected temperatures for $C_0$, $C_2^I$ and $C_2^{II}$ respectively. The slopes (relatively constant for different temperatures measured) of the linear fitting determine the Dingle temperatures $T_D$, which are $15\ K$ for $C_0$, $3\ K$ for $C_2^I$ and $40\ K$ for $C_2^{II}$, corresponding to quantum (SdHO) mobilities ($\mu_s^{SdH} = \frac{e\hbar}{2\pi T_D k_B m_c}$) $\sim 1800\ cm^2/Vs$, $2000\ cm^2/Vs$ and $300\ cm^2/Vs$, respectively.

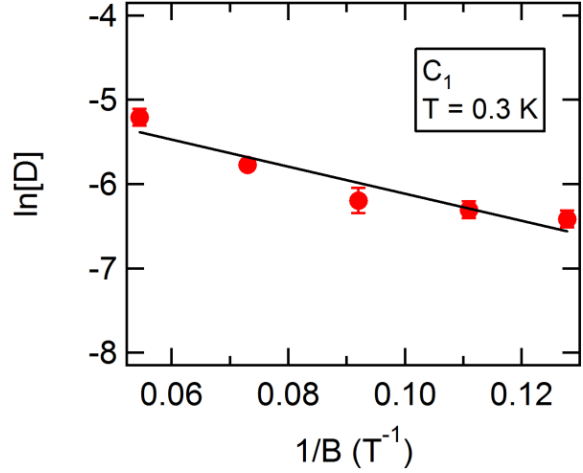

**Figure S9**. Dingle analysis performed to extract the quantum mobility ($\mu_s^{SdH}$) from the SdHO in $C_1$ (measured only at T = 0.3 K, thus we cannot perform a full Lifshitz-Kosevich analysis as those in Fig. S8a and b). Given the fact that SdHO frequency of $C_1$ is close to $C_2^I$ (indicating similar Fermi levels), we assume $m_c$ (of the 2DEG) of sample $C_1$ is comparable with that extracted from $C_2^I$. From the linear fitting (the black solid line) in the Dingle plot, we extract the Dingle temperature of $C_1$ to be $T_D \sim 3$ K, corresponding to an SdHO mobility $\mu_s^{SdH} \sim 2000\ cm^2/Vs$.

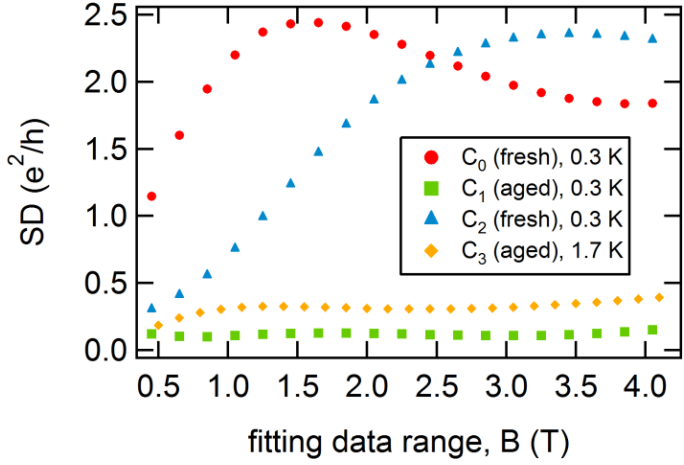

**Figure S10.** To evaluate our fitting of WAL behavior with the HLN theory, a standard deviation ($SD$) was defined as $SD(N) = \sqrt{\frac{1}{N}\sum_{i=1}^{N}[\Delta G_{exp}(B_i) - \Delta G_{HLN}(B_i)]^2}$, where $\Delta G_{exp}(B_i)$ is the i$^{th}$ experimental data point and $\Delta G_{HLN}(B_i)$ is calculated by the HLN formula with the fitting parameters obtained from a fitting range $[0, B_N]$. We plot $SD$ as a function of $B_N$. The standard deviations of $C_1$ and $C_3$ (both aged) are: 1) much smaller than those of $C_0$ and $C_2$ (both fresh-cleaved), and 2) similar for different fitting ranges. This indicates HLN theory can capture the transport feature (WAL) measured in an aged sample to a better degree than in a fresh-cleaved one.

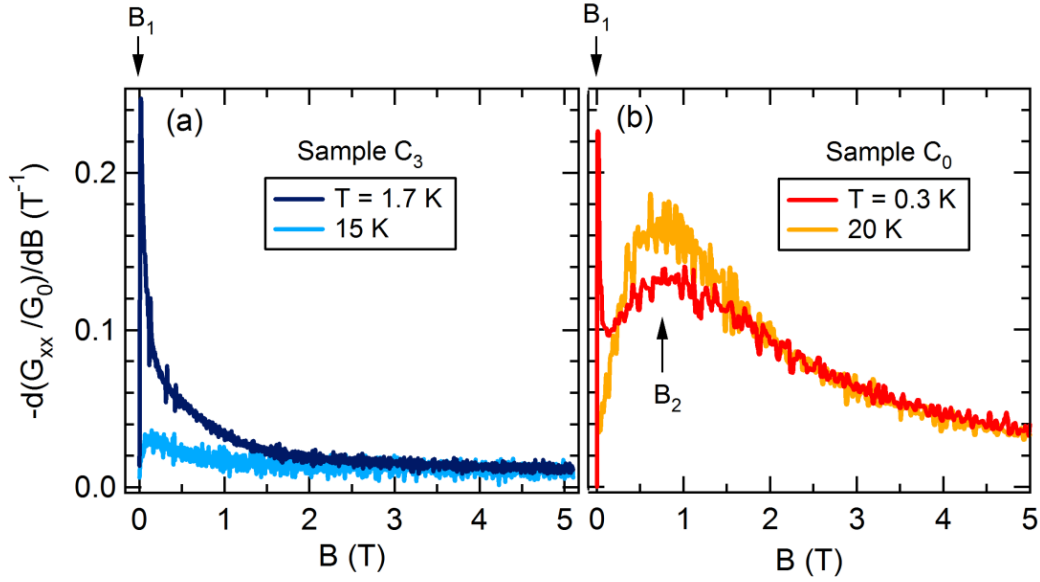

**Figure S11.** (a, b) Derivatives of the normalized magneto-conductance, $\frac{-dG_{xx}/G_0}{dB}$, as functions of B measured in sample $C_3$ and $C_0$, respectively, at different temperatures (T). The sharp peak at $B_1$ corresponding to HLN theory for WAL (see the main text) is found to decrease with elevated T for both samples, as expected for a quantum coherent transport and consistent with the typical WAL behavior. On the contrary, the peak at $B_2$ observed in sample $C_0$ is noted to *increase* with increasing T (opposite to the typical WAL behavior). More theoretical and experimental studies are needed to fully understand this feature.

**Table S1.** Surface conditions of sample C, listed in chronological order from $C_0$ to $C_2$. In this table, $t_{int}$ denotes the total amount of time lapse since the initial measurements on $C_0$, when a sample was cooled down.

| Sample | $t_{int}$ (days) | surface condition | exposure time (since the last fresh cleavage) |
|---|---|---|---|
| $C_0$ | - | freshly cleaved | < 30 seconds |
| $C_0'$ | 21 | aged | 20 days |
| $C_1$ | 74 | aged | 73 days |
| $C_2$ | 76 | freshly cleaved again | ~ 2 mins |
| $C_3$ | 90 | aged | 13 days |

**Table S2.** Parameters used to calculate the WAL theoretical (HLN) curves shown in Fig. 4d.

| Sample | $l_\phi$ (nm) | $-\alpha$ |
|---|---|---|
| $C_0$ | 190 | 22 |
| $C_1$ | 120 | 22 |
| $C_2$ | 105 | 30 |
| $C_3$ | 135 | 13 |

**Supplementary Reference**